\catcode`\@=11					



\font\fiverm=cmr5				
\font\fivemi=cmmi5				
\font\fivesy=cmsy5				
\font\fivebf=cmbx5				

\skewchar\fivemi='177
\skewchar\fivesy='60


\font\sixrm=cmr6				
\font\sixi=cmmi6				
\font\sixsy=cmsy6				
\font\sixbf=cmbx6				

\skewchar\sixi='177
\skewchar\sixsy='60


\font\sevenrm=cmr7				
\font\seveni=cmmi7				
\font\sevensy=cmsy7				
\font\sevenit=cmti7				
\font\sevenbf=cmbx7				

\skewchar\seveni='177
\skewchar\sevensy='60


\font\eightrm=cmr8				
\font\eighti=cmmi8				
\font\eightsy=cmsy8				
\font\eightit=cmti8				
\font\eightbf=cmbx8				

\skewchar\eighti='177
\skewchar\eightsy='60


\font\ninei=cmmi9
\font\ninesy=cmsy9

\skewchar\ninei='177
\skewchar\ninesy='60


\font\tenrm=cmr10				
\font\teni=cmmi10				
\font\tensy=cmsy10				
\font\tenex=cmex10				
\font\tenit=cmti10				
\font\tensl=cmsl10				
\font\tenbf=cmbx10				
\font\tentt=cmtt10				
\font\tenss=cmss10				
\font\tensc=cmcsc10				
\font\tenbi=cmmib10				

\skewchar\teni='177
\skewchar\tenbi='177
\skewchar\tensy='60

\def\tenpoint{\ifmmode\err@badsizechange\else
	\textfont0=\tenrm \scriptfont0=\sevenrm \scriptscriptfont0=\fiverm
	\textfont1=\teni  \scriptfont1=\seveni  \scriptscriptfont1=\fivemi
	\textfont2=\tensy \scriptfont2=\sevensy \scriptscriptfont2=\fivesy
	\textfont3=\tenex \scriptfont3=\tenex   \scriptscriptfont3=\tenex
	\textfont4=\tenit \scriptfont4=\sevenit \scriptscriptfont4=\sevenit
	\textfont5=\tensl
	\textfont6=\tenbf \scriptfont6=\sevenbf \scriptscriptfont6=\fivebf
	\textfont7=\tentt
	\textfont8=\tenbi \scriptfont8=\seveni  \scriptscriptfont8=\fivemi
	\def\rm{\tenrm\fam=0 }%
	\def\it{\tenit\fam=4 }%
	\def\sl{\tensl\fam=5 }%
	\def\bf{\tenbf\fam=6 }%
	\def\tt{\tentt\fam=7 }%
	\def\ss{\tenss}%
	\def\sc{\tensc}%
	\def\bmit{\fam=8 }%
	\rm\setparameters\setbaselines\fi}


\font\twelverm=cmr12				
\font\twelvei=cmmi12				
\font\twelvesy=cmsy10	scaled\magstep1		
\font\twelveex=cmex10	scaled\magstep1		
\font\twelveit=cmti12				
\font\twelvesl=cmsl12				
\font\twelvebf=cmbx12				
\font\twelvett=cmtt12				
\font\twelvess=cmss12				
\font\twelvesc=cmcsc10	scaled\magstep1		
\font\twelvebi=cmmib10	scaled\magstep1		

\skewchar\twelvei='177
\skewchar\twelvebi='177
\skewchar\twelvesy='60

\def\twelvepoint{\ifmmode\err@badsizechange\else
	\textfont0=\twelverm \scriptfont0=\eightrm \scriptscriptfont0=\sixrm
	\textfont1=\twelvei  \scriptfont1=\eighti  \scriptscriptfont1=\sixi
	\textfont2=\twelvesy \scriptfont2=\eightsy \scriptscriptfont2=\sixsy
	\textfont3=\twelveex \scriptfont3=\tenex   \scriptscriptfont3=\tenex
	\textfont4=\twelveit \scriptfont4=\eightit \scriptscriptfont4=\sevenit
	\textfont5=\twelvesl
	\textfont6=\twelvebf \scriptfont6=\eightbf \scriptscriptfont6=\sixbf
	\textfont7=\twelvett
	\textfont8=\twelvebi \scriptfont8=\eighti  \scriptscriptfont8=\sixi
	\def\rm{\twelverm\fam=0 }%
	\def\it{\twelveit\fam=4 }%
	\def\sl{\twelvesl\fam=5 }%
	\def\bf{\twelvebf\fam=6 }%
	\def\tt{\twelvett\fam=7 }%
	\def\ss{\twelvess}%
	\def\sc{\twelvesc}%
	\def\bmit{\fam=8 }%
	\rm\setparameters\setbaselines\fi}


\font\fourteenrm=cmr12	scaled\magstep1		
\font\fourteeni=cmmi12	scaled\magstep1		
\font\fourteensy=cmsy10	scaled\magstep2		
\font\fourteenex=cmex10	scaled\magstep2		
\font\fourteenit=cmti12	scaled\magstep1		
\font\fourteensl=cmsl12	scaled\magstep1		
\font\fourteenbf=cmbx12	scaled\magstep1		
\font\fourteentt=cmtt12	scaled\magstep1		
\font\fourteenss=cmss12	scaled\magstep1		
\font\fourteensc=cmcsc10 scaled\magstep2	
\font\fourteenbi=cmmib10 scaled\magstep2	

\skewchar\fourteeni='177
\skewchar\fourteenbi='177
\skewchar\fourteensy='60

\def\fourteenpoint{\ifmmode\err@badsizechange\else
	\textfont0=\fourteenrm \scriptfont0=\tenrm \scriptscriptfont0=\sevenrm
	\textfont1=\fourteeni  \scriptfont1=\teni  \scriptscriptfont1=\seveni
	\textfont2=\fourteensy \scriptfont2=\tensy \scriptscriptfont2=\sevensy
	\textfont3=\fourteenex \scriptfont3=\tenex \scriptscriptfont3=\tenex
	\textfont4=\fourteenit \scriptfont4=\tenit \scriptscriptfont4=\sevenit
	\textfont5=\fourteensl
	\textfont6=\fourteenbf \scriptfont6=\tenbf \scriptscriptfont6=\sevenbf
	\textfont7=\fourteentt
	\textfont8=\fourteenbi \scriptfont8=\tenbi \scriptscriptfont8=\seveni
	\def\rm{\fourteenrm\fam=0 }%
	\def\it{\fourteenit\fam=4 }%
	\def\sl{\fourteensl\fam=5 }%
	\def\bf{\fourteenbf\fam=6 }%
	\def\tt{\fourteentt\fam=7}%
	\def\ss{\fourteenss}%
	\def\sc{\fourteensc}%
	\def\bmit{\fam=8 }%
	\rm\setparameters\setbaselines\fi}


\font\seventeenrm=cmr10 scaled\magstep3		


\newdimen\rp@
\newcount\@basestretchnum
\newskip\@baseskip
\newskip\headskip
\newskip\footskip


\def\setparameters{\rp@=.1em
	\headskip=24\rp@
	\footskip=\headskip
	\delimitershortfall=5\rp@
	\nulldelimiterspace=1.2\rp@
	\scriptspace=0.5\rp@
	\abovedisplayskip=10\rp@ plus3\rp@ minus5\rp@
	\belowdisplayskip=10\rp@ plus3\rp@ minus5\rp@
	\abovedisplayshortskip=5\rp@ plus2\rp@ minus4\rp@
	\belowdisplayshortskip=10\rp@ plus3\rp@ minus5\rp@
	\normallineskip=\rp@
	\lineskip=\normallineskip
	\normallineskiplimit=0pt
	\lineskiplimit=\normallineskiplimit
	\jot=3\rp@
	\setbox0=\hbox{\the\textfont3 B}\p@renwd=\wd0
	\skip\footins=12\rp@ plus3\rp@ minus3\rp@
	\skip\topins=0pt plus0pt minus0pt}


\def\setbaselines{\maxdepth=4\rp@\baselinestretch=\@basestretchnum}


\def\baselinestretch{\afterassignment\@basestretch\@basestretchnum}
\def\@basestretch{%
	\@baseskip=12\rp@ \divide\@baseskip by1000
	\normalbaselineskip=\@basestretchnum\@baseskip
	\baselineskip=\normalbaselineskip
	\bigskipamount=\the\baselineskip
		plus.25\baselineskip minus.25\baselineskip
	\medskipamount=.5\baselineskip
		plus.125\baselineskip minus.125\baselineskip
	\smallskipamount=.25\baselineskip
		plus.0625\baselineskip minus.0625\baselineskip
	\setbox\strutbox=\hbox{\vrule height.708\baselineskip
		depth.292\baselineskip width0pt }}



\def\makeheadline{\vbox to0pt{\baselinestretch=1000
	\vskip-\headskip \vskip1.5pt
	\line{\vbox to\ht\strutbox{}\the\headline}\vss}\nointerlineskip}

\def\makefootline{\baselineskip=\footskip\line{\the\footline}}

\def\big#1{{\hbox{$\left#1\vbox to8.5\rp@ {}\right.\n@space$}}}
\def\Big#1{{\hbox{$\left#1\vbox to11.5\rp@ {}\right.\n@space$}}}
\def\bigg#1{{\hbox{$\left#1\vbox to14.5\rp@ {}\right.\n@space$}}}
\def\Bigg#1{{\hbox{$\left#1\vbox to17.5\rp@ {}\right.\n@space$}}}


\mathchardef\alpha="710B
\mathchardef\beta="710C
\mathchardef\gamma="710D
\mathchardef\delta="710E
\mathchardef\epsilon="710F
\mathchardef\zeta="7110
\mathchardef\eta="7111
\mathchardef\theta="7112
\mathchardef\iota="7113
\mathchardef\kappa="7114
\mathchardef\lambda="7115
\mathchardef\mu="7116
\mathchardef\nu="7117
\mathchardef\xi="7118
\mathchardef\pi="7119
\mathchardef\rho="711A
\mathchardef\sigma="711B
\mathchardef\tau="711C
\mathchardef\upsilon="711D
\mathchardef\phi="711E
\mathchardef\chi="711F
\mathchardef\psi="7120
\mathchardef\omega="7121
\mathchardef\varepsilon="7122
\mathchardef\vartheta="7123
\mathchardef\varpi="7124
\mathchardef\varrho="7125
\mathchardef\varsigma="7126
\mathchardef\varphi="7127
\mathchardef\imath="717B
\mathchardef\jmath="717C
\mathchardef\ell="7160
\mathchardef\wp="717D
\mathchardef\partial="7140
\mathchardef\flat="715B
\mathchardef\natural="715C
\mathchardef\sharp="715D


\def\err@badsizechange{%
	\immediate\write16{--> Size change not allowed in math mode, ignored}}

\baselinestretch=1000
\tenpoint

\catcode`\@=12					
\catcode`\@=11
\expandafter\ifx\csname @iasmacros\endcsname\relax
	\global\let\@iasmacros=\par
\else	\immediate\write16{}
	\immediate\write16{Warning:}
	\immediate\write16{You have tried to input iasmacros more than once.}
	\immediate\write16{}
	\endinput
\fi
\catcode`\@=12


\def\rmb{\seventeenrm}

\def\singlespace{\baselineskip=\normalbaselineskip}
\def\halfspace{\baselineskip=1.5\normalbaselineskip}
\def\doublespace{\baselineskip=2\normalbaselineskip}


\def\AB{\bigskip\parindent=40pt
        \centerline{\bf ABSTRACT}\medskip\halfspace\narrower}
\def\AE{\bigskip\nonarrower\doublespace}
\def\nonarrower{\advance\leftskip by-\parindent
	\advance\rightskip by-\parindent}


\def\boxit#1{\vbox{\hrule\hbox{\vrule\kern3pt
	\vbox{\kern3pt#1\kern3pt}\kern3pt\vrule}\hrule}}

\def\hence{\leavevmode\hbox{\bf .\raise5.5pt\hbox{.}.} }

\def\dalemb#1#2{{\vbox{\hrule height.#2pt
	\hbox{\vrule width.#2pt height#1pt \kern#1pt \vrule width.#2pt}
	\hrule height.#2pt}}}
\def\gtorder{\mathrel{\raise.3ex\hbox{$>$}\mkern-14mu
             \lower0.6ex\hbox{$\sim$}}}
\def\ltorder{\mathrel{\raise.3ex\hbox{$<$}\mkern-14mu
             \lower0.6ex\hbox{$\sim$}}}

\newdimen\fullhsize
\newbox\leftcolumn
\def\twoup{\hoffset=-.5in \voffset=-.25in
  \hsize=4.75in \fullhsize=10in \vsize=6.9in
  \def\fullline{\hbox to\fullhsize}
  \let\lr=L
  \output={\if L\lr
        \global\setbox\leftcolumn=\columnbox\global\let\lr=R \advancepageno
      \else \doubleformat \global\let\lr=L\fi
    \ifnum\outputpenalty>-20000 \else\dosupereject\fi}
  \def\doubleformat{\shipout\vbox{
    \fullline{\box\leftcolumn\hfil\columnbox}\advancepageno}}
  \def\columnbox{\leftline{\vbox{\makeheadline\pagebody\makefootline}}}
  \tolerance=1000 }

\twelvepoint
\doublespace
\overfullrule=0pt
{\nopagenumbers{
\rightline{~~~November, 2001}
\bigskip\bigskip
\centerline{\rmb  Environmental Influence}
\centerline{\rmb on the Measurement Process in Stochastic Reduction Models}
\medskip
\centerline{\it Stephen L. Adler
}
\centerline{\bf Institute for Advanced Study}
\centerline{\bf Princeton, NJ 08540}
\medskip
\bigskip\bigskip
\leftline{\it Send correspondence to:}
\medskip
{\singlespace\leftline{Stephen L. Adler}
\leftline{Institute for Advanced Study}
\leftline{Einstein Drive, Princeton, NJ 08540}
\leftline{Phone 609-734-8051; FAX 609-924-8399; email adler@ias.
edu}}
\bigskip\bigskip
}}
\vfill\eject
\pageno=2
\AB
We consider the energy-driven stochastic state vector 
reduction equation for the density matrix, which for pure state density 
matrices can be written in two equivalent forms.   We use these forms 
to discuss the decoupling of the noise terms for 
independent subsystems,  and to construct ``environmental'' 
stochastic density matrices whose time-independent expectations are the usual 
quantum statistical distributions. We then consider a  
measurement apparatus weakly coupled to an external environment, and show 
that in mean field (Hartree) approximation the stochastic equation separates 
into independent equations for the apparatus and environment, with the 
Hamiltonian for the apparatus augmented by the environmental  
expectation of the interaction Hamiltonian.  We use the Hartree approximated 
equation to study a simple accretion model for the interaction of the  
apparatus with its environment, as part of a more general discussion of   
when the stochastic dynamics predicts state vector reduction, and when it   
predicts the maintenance of coherence.  We also discuss the magnitude of 
decoherence effects acting during the reduction process. 
Our analysis supports the suggestion that 
a measurement takes place when the different outcomes 
are characterized by sufficiently distinct environmental interactions for 
the reduction process to be rapidly driven to completion.  
\AE
\bigskip\bigskip
\vfill\eject
\pageno=3
\centerline{{\bf 1.~~Introduction}}
Understanding the measurement process has been a persistent problem since 
the inception of quantum mechanics.  In the orthodox Copenhagen 
interpretation, measurements are accounted for by invoking a layer of 
classical, non-quantum mechanical reality; attempts to extend quantum 
mechanics to include the measuring apparatus itself lead to quandaries, 
such as the famous Schr\"odinger cat paradox.  One approach to this 
problem that has been much studied recently [1-8] postulates that the 
Schr\"odinger equation is only an approximate description of reality,    
and must be modified by small, nonlinear stochastic terms.  
These terms drive the state vector reduction process,  
and account for the non-observation of macroscopic quantum superpositions 
in measurement situations.  

The proposal that a stochastic, nonlinear Schr\"odinger equation 
provides the phenomenology of quantum measurement passes a number of 
consistency tests.  In its energy-driven form, it leads 
exactly to probabilities given by the Born rule [4,8,9], and 
for measurements on  
degenerate systems leads to the L\"uders projection rule [9].  There are 
plausible arguments [7,8], to be elaborated on here,  
that with a Planckian magnitude 
of the stochastic term, 
coherence is maintained where observed experimentally, while state 
vector reduction proceeds for measurement situations where discrete     
outcomes are observed.  
However, the stochastic Schr\"odinger equation is inherently 
nonrelativistic [10,11], 
involving the same stochastic differential at all spatial locations.  This 
raises the issue of whether it is consistent with clustering -- put simply, 
does the reduction of the state vector in a 
localized measuring apparatus proceed independently of 
what goes on far away from the laboratory?  An affirmative answer to this 
question was given [8] under the  assumption that all of the universe 
is governed by the pure state stochastic reduction equation.   
In this paper we extend this analysis in a number of directions, 
with the aim of understanding in greater detail 
the stochastic evolution of a ``measurement'' system coupled to 
its environment.  

Our discussion is organized as follows.  In Sec. 2 we give two 
equivalent forms of the It\^o noise term in 
the stochastic evolution equation for a pure state 
density matrix, and use these to discuss clustering for disjoint subsystems.  
In Sec. 3 we use one of these forms to prove the existence of pure state 
density matrices whose stochastic expectation gives the standard 
quantum statistical distributions.  We also give a mixed state generalization 
of these results that is relevant when the Hamiltonian has degeneracies.  
In Sec. 4 we consider 
a ``measurement'' subsystem  
weakly coupled to an ``environment'' subsystem, obeying overall  
the density matrix stochastic evolution equation, and 
derive the corresponding single system Hartree or mean field stochastic 
equations for the measurement and environment subsystems, 
working to first order accuracy  in the interaction Hamiltonian.

In Sec. 5 we give a survey of under what circumstances 
the stochastic evolution equation predicts state vector reduction, based on   
Planckian estimates for the magnitude of the stochastic term.  We show 
that when the energy spread between superimposed states is small, coherence 
is maintained, in agreement with recent experiments on  
quantum coherence in large systems.  On the other hand, when the energy 
fluctuations in a measurement system are large enough, state vector  
reduction proceeds rapidly to completion.  We consider three types of 
energy fluctuations:  thermal fluctuations, shot effect fluctuations in 
electric currents, and surface accretion fluctuations.  For the latter, 
we use the mean field approximation of Sec. 4 to construct 
a simple model for accretion processes, which motivates a  
quantitative discussion of their influence on the measurement process in  
both terrestrial and extraterrestrial environments.  
In Sec. 6 we discuss the coexistence of standard decoherence mechanisms 
with the stochastic reduction process.  
In Sec. 7 we state our conclusions regarding 
the implications of this analysis for the 
measurement process in quantum mechanics.  In an Appendix, we discuss a 
coherent state variant of the accretion model of Sec. 5, in which 
reduction can proceed to coherent states.  
\bigskip

\centerline{{\bf 2.~~Stochastic Density Matrix Equations and Clustering}}
\bigskip
We begin by recalling some formulas from the theory of 
stochastic Schr\"odinger equations [1-8].  Letting $|\chi\rangle$ be a 
normalized state vector, the standard stochastic evolution (``quantum 
state diffusion'') equation for $|\chi\rangle$ takes the form
$$d|\chi\rangle=[\alpha dt + \beta dW_t] |\chi \rangle~~~,  \eqno(1a)$$
with $dW_t$ a real It\^o stochastic differential obeying 
$$dW_t^2=dt~,~~~dW_t dt=0~~~,\eqno(1b)$$
and with 
$$\eqalign{
\alpha=&-iH-{1\over 8} \sigma^2 [A-\langle A \rangle]^2~~~,\cr
\beta=&{1\over 2} \sigma [A-\langle A \rangle]~~~,\cr
}\eqno(1c)$$
where $\sigma$ is a numerical parameter governing the strength of the 
stochastic and drift terms, and $A$ is a self-adjoint 
operator with expectation 
$\langle A \rangle$ in the state $|\chi\rangle$, 
$$\langle A \rangle = \langle \chi|A|\chi \rangle~~~. \eqno(1d)$$
The energy-driven case of the stochastic evolution is obtained by taking 
$A$ to be the Hamiltonian $H$, which we shall assume henceforth.  It is 
straightforward to show that the evolution of Eqs.(1a)--(1d) preserves the 
normalization of the state vector $|\chi\rangle$.

Defining the pure state density matrix $\rho=|\chi\rangle \langle \chi|$, 
it is easy to show that the state vector evolution of Eqs.~(1a)--(1d) 
implies that the density matrix evolution is given by  
$$d\rho=-i[H,\rho] dt - {1\over 8}\sigma^2 [H,[H,\rho]]dt   
+{1 \over 2} \sigma N(\rho,H) dW_t ~~~. \eqno(2)$$
Direct calculation from Eqs.~(1a)--(1d) gives
the coefficient $N(\rho,H)$ of the It\^o noise term $dW_t$ in Eq.~(2) 
in the form 
$$N(\rho,H)=\{\rho,H\}-2\rho {\rm Tr}\rho H~~~~,\eqno(3a)$$ 
which by use of the pure state condition $\rho^2=\rho$  
can be written in the equivalent form 
$$N(\rho,H)= [\rho,[\rho,H]]~~~~.\eqno(3b)$$   
Both of these forms have the property that $\rho^2=\rho$ implies 
that $\{\rho,d\rho\}+(d\rho)^2=d\rho$, which can be rewritten as 
$(\rho+d\rho)^2=\rho+d\rho$, and so they preserve the pure state condition.

Let us now consider a system for which the Hamiltonian 
$H$ is the sum of two Hamiltonians $H_1~, H_2$ which depend on disjoint 
sets of variables, and investigate the conditions under which Eqs.~(2) 
and (3a), (3b) 
admit factorized solutions $\rho=\rho_1\rho_2$,  with $\rho_{1,2}$ 
obeying equations of similar form driven by the respective Hamiltonians 
$H_{1,2}$.  Substituting $H=H_1+H_2$ and $\rho=\rho_1\rho_2$ 
into Eqs.~(3a), (3b),  
and using the facts that all variables in set 1 commute with all 
variables in set 2, and that ${\rm Tr}={\rm Tr}_1 {\rm Tr}_2$, we 
find respectively from Eqs.~(3a) and (3b) that
$$N(\rho_1\rho_2,H_1+H_2)= 
\rho_2[\{\rho_1,H_1\}-2\rho_1 {\rm Tr_2} \rho_2 {\rm Tr_1}\rho_1 H_1] 
+\rho_1[\{\rho_2,H_2\}-2\rho_2 {\rm Tr_1} \rho_1 {\rm Tr_2}\rho_2 H_2] 
~~~~,\eqno(4a)$$
$$N(\rho_1\rho_2,H_1+H_2)= 
 \rho_2^2 [\rho_1,[\rho_1,H_1]]  + \rho_1^2 [\rho_2,[\rho_2,H_2]] 
~~~.~~~~~~~~~~~~~~~~~~~~~~~~~~~~~~~~~~~~~~~~~~\eqno(4b)$$  
Clustering requires that 
$$N(\rho_1\rho_2,H_1+H_2)=\rho_2 N_1(\rho_1,H_1) + \rho_1 N_2(\rho_2,H_2)
~~~,\eqno(5)$$
with $N_{1,2}$ the restrictions of $N$ to the 1,2 subspaces.  
We see that Eq.~(4a) obeys the clustering property by virtue of the 
trace conditions ${\rm Tr_1}\rho_1=1$, ${\rm Tr_2}\rho_2=1$, while 
Eq.~(4b) satisfies the clustering property by virtue of the pure state 
conditions $\rho_1^2=\rho_1$, $\rho_2^2=\rho_2$.   

Let us now examine the clustering properties of the remaining terms in 
Eq.~(2).  For the left hand side, we find by use of the It\^o extension 
of the chain rule that
$$d(\rho_1\rho_2)=\rho_2 d\rho_1 +\rho_1 d\rho_2 +d\rho_1d\rho_2
~~~.\eqno(6a)$$
Thus, in order to have $d\rho_1$ and $d\rho_2$ obeying equations of the 
same form as $d\rho$ but restricted to the $1,2$ subspaces, the left hand 
side should take the form, using Eqs.~(1b) and (2), 
$$d(\rho_1\rho_2)=\rho_2 d\rho_1 +\rho_1 d\rho_2 + {1\over 4} \sigma^2 
N_1(\rho_1,H_1)
N_2(\rho_2,H_2) dt
~~~.\eqno(6b)$$
{}For the $dt$ terms on the right hand side of Eq.~(2), we have   
$$\eqalign{
-i[H_1+H_2,\rho_1\rho_2] dt -&{1\over 8}\sigma^2 [H_1+H_2,[H_1+H_2,
\rho_1\rho_2]]dt\cr   
=&\rho_2\{-i[H_1,\rho_1]dt -{1\over 8}\sigma^2 [H_1,[H_1,\rho_1]] dt \}\cr
+&\rho_1\{-i[H_2,\rho_2]dt -{1\over 8}\sigma^2 [H_2,[H_2,\rho_2]] dt \}\cr
-&{1\over 4}\sigma^2 [H_1,\rho_1][H_2,\rho_2] dt~~~.\cr 
}\eqno(6c)$$
Assuming the conditions for the clustering property of Eq.~(5) to hold for 
the It\^o noise term, comparing Eqs.~(6a)-(6c) we see that 
the complete density matrix evolution equation will 
cluster if and only if 
$$N_1(\rho_1,H_1) N_2(\rho_2,H_2)
= -[H_1,\rho_1][H_2,\rho_2] ~~~.\eqno(7)$$  
This condition does not hold as in identity for either of the two possible 
forms for $N(\rho,H)$ given in Eqs.~(3a), (3b), and so the $\sigma^2 dt$ or   
drift term in the stochastic evolution equation  does couple disjoint  
systems.  

However, there are two important special cases in which disjoint 
systems decouple asymptotically.  The first of these cases corresponds 
[8] to taking $N(\rho,H)$ as in Eq.~(3b), so that 
Eq.~(7) becomes
$$ [\rho_1,[\rho_1,H_1]] [\rho_2,[\rho_2,H_2]]
= -[H_1,\rho_1][H_2,\rho_2] ~~~.\eqno(8a)$$  
This equation is satisfied, by virtue of both the left and right hand sides 
vanishing, whenever either $[\rho_1,H_1]=0$ or $[\rho_2,H_2]=0$, conditions  
that are respectively obeyed when system 1 or system 2 is at the 
endpoint of the state vector reduction process.  In particular, if system 
1 represents a measurement process, and system 2 represents a pure state 
environment at the endpoint of its reduction process, then the stochastic 
dynamics of system 1 is completely independent of the dynamics of its 
environment.   

A more general case in which disjoint systems decouple 
asymptotically corresponds to taking $N(\rho,H)$ as in Eq.~(3a), but not 
assuming the pure state condition so that this cannot be transformed to 
Eq.~(3b).  Equation (7) now becomes 
$$[\{\rho_1,H_1\}-2\rho_1{\rm Tr_1}\rho_1 H_1] 
[\{\rho_2,H_2\}-2\rho_2{\rm Tr_2}\rho_2 H_2]= 
-[H_1,\rho_1][H_2,\rho_2]~~~~.\eqno(8b)$$
This equation is satisfied, again by virtue of both the left and right hand 
sides vanishing, whenever either $\rho_1$ is a linear combination of 
projectors on a degenerate 
submanifold of $H_1$, or $\rho_2$ is a linear combination of  
projectors on a degenerate submanifold 
of $H_2$.  For example, in the latter case we would have $\rho_2 H_2=
H_2 \rho_2=E_2 \rho_2$ for some degenerate submanifold energy $E_2$, together 
with ${\rm Tr_2}\rho_2=1$, which imply the simultaneous vanishing of 
$\{\rho_2,H_2\}-2\rho_2{\rm Tr_2}\rho_2 H_2 $ and of $[H_2,\rho_2]$.  Thus, 
if one were to adopt Eqs.~(2) and (3a) as a generalization of the density 
matrix evolution equation to the case of non-pure state density matrices, 
a pure state measurement process decouples from a mixed state environment 
whenever the density matrix for this environment is a linear combination of 
projectors on a degenerate submanifold of its Hamiltonian.  

\bigskip
\centerline{\bf 3.~~Martingale Construction of the Standard} 
\centerline{\bf Quantum Statistical Distributions}
\bigskip

In order for the measurement system to decouple from its environment, we
have seen that the environment must be described either 
by a pure state density matrix that commutes with 
the environment Hamiltonian, or by   
a mixed state density matrix that is a linear combination of projectors 
on a degenerate submanifold of 
the environment Hamiltonian (with the second case equivalent to the 
first  for a one dimensional submanifold).  
This raises the question of how such a description can be compatible 
with the usual description of equilibrium environments in terms of the 
standard quantum statistical distributions, which are mixed state density 
matrices $\rho$ obeying the trace condition ${\rm Tr}\rho=1$, 
but which do not obey 
either the pure state condition $\rho^2=\rho$ or the more general 
condition that the density matrix be a linear combination of projectors 
on a degenerate Hamiltonian 
submanifold.  The answer is that in the theory of 
stochastic state vector reduction, the role of the usual mixed state 
density matrix is played [7] by the stochastic expectation $E[\rho]$  
and not by $\rho$ itself.  Thus 
an equilibrium environment can be described by a stochastic density 
matrix that is a 
linear combination of projectors on a degenerate Hamiltonian submanifold, 
the stochastic expectation of which has the form
$E[\rho]=f(H)$, with $f$ one of the standard quantum statistical 
distribution functions of the Hamiltonian.  Since Eq.~(2) implies that 
$E[\rho]$ obeys the time evolution equation 
$$dE[\rho]=-i[H,E[\rho]]dt-{1\over 8} 
\sigma^2 [H,[H,E[\rho]]]dt~~~,\eqno(9)$$ 
any $E[\rho]$ of the form $f(H)$ is time independent, as expected of the 
quantum statistical distributions.  

To show that there are pure state density matrices with the required 
expectation, we proceed constructively by use of the density matrix 
evolution equation in the form 
$$d\rho=-i[H,\rho] dt - {1\over 8}\sigma^2 [H,[H,\rho]]dt   
+{1 \over 2} \sigma[\{\rho,H\}-2\rho {\rm Tr}\rho H] dW_t ~~~. \eqno(10)$$
Although we derived this equation in Sec. 2 for pure state density matrices,  
we shall now use it, as suggested in the discussion associated with Eq.~(8b),  
as a stochastic evolution equation for density 
matrices $\rho$  that do not obey the pure state condition.
Taking the initial $\rho$ at time $t=0$ as $\rho_0=f(H)$, we see from 
Eq.~(9), which follows by taking the expectation of Eq.~(10), that 
$E[\rho]=f(H)$ for all times.  Also, since Eq.~(10) only involves 
the Hamiltonian H, the stochastically evolved $\rho$ is still a function 
of $H$, and so commutes with $H$ at all times.  
Thus, for the choice of initial condition $\rho_0=f(H)$, Eq.~(10) simplifies
to 
$$d\rho={1 \over 2} \sigma[\{\rho,H\}-2\rho {\rm Tr}\rho H] dW_t ~~~. 
\eqno(11)$$
Equation (11) defines $\rho$ to be a martingale, for which the expectation 
$E_s$ conditional on information available up to time $s$ obeys 
$E_s[\rho_t]=\rho_s~,~~s \leq t$, which reduces to $E[\rho]\equiv E_0[\rho_t]
=\rho_0=f(H)$ when $s=0$.  [Note that if instead of Eq.~(10) we had used the 
stochastic equation obtained from Eqs.~(2) and (3b), the initial 
$\rho_0=f(H)$ would not evolve in time at all, since Eq.~(3b) vanishes 
identically when $\rho$ commutes with $H$.  This underscores again the fact 
that Eqs.~(3a) and (3b) are equivalent only for pure state density matrices, 
but define different stochastic evolutions for density matrices not obeying 
the pure state condition $\rho^2=\rho$.]

Let us now show that at late times  
$\rho$ evolves by Eq.~(11) into a pure state projector 
when the Hamiltonian $H$ is   
nondegenerate, or into a linear combination of projectors on a degenerate 
submanifold of $H$ when $H$ is degenerate.  
The proof of this parallels the 
proof [8,9] that Eqs.~(1a)--(1d) lead to state vector reduction.  We consider 
the variance $V$ of the Hamiltonian, defined by 
$$V={\rm Tr}\rho H^2 -({\rm Tr} \rho H)^2~~~,\eqno(12a)$$ 
which by the It\^o extension of the chain rule evolves in time as 
$$dV={\rm Tr} d\rho H^2 - 2{\rm Tr} \rho H {\rm Tr}d\rho H
-({\rm Tr}d\rho H)^2~~~.\eqno(12b)$$
Using  Eq.~(11) for $d\rho$ to evaluate ${\rm Tr}d\rho H^n$, we find  
$${\rm Tr}d\rho H^n=\sigma [{\rm Tr} \rho H^{n+1}-{\rm Tr} 
\rho H^n {\rm Tr}\rho H]dW_t ~~~.\eqno(12c)$$
Thus, substituting Eq.~(12c) for $n=1,~2$ into Eq.~(12b) 
and taking the expectation, 
we get 
$$dE[V]=-\sigma^2 E[V^2]dt~~~.\eqno(13)$$
{}From here on the argument is identical to that of Refs. [8,9], and leads 
to the conclusion that as $t \to \infty$ the variance $V$ approaches 0 
almost certainly.  When the energy spectrum is nondegenerate, this implies 
that at late times 
only one density matrix element $\rho_E$ is nonzero, 
and so the initial density matrix $\rho_0=f(H)$ has evolved to a pure state 
density matrix obeying $\rho^2=\rho$.    
More generally,  when the energy spectrum is degenerate, the 
vanishing of the variance implies that the density matrix has evolved to 
a linear combination of projectors on a degenerate submanifold of the 
Hamiltonian.  
Thus, evolution of the initial density matrix $\rho_0=f(H)$ by Eq.~(10)  
leads to a late time density matrix that obeys  
$E[\rho]=f(H)$, and which is a a pure state density matrix in the 
nondegenerate case, or a linear combination of projectors on a degenerate 
submanifold of $H$ in the degenerate case.  
We take such density matrices as our model for the 
environment, and by the arguments of Sec.~2, are assured that the evolution 
of measurement systems uncoupled by Hamiltonian interaction terms to this 
environment are independent of the environmental dynamics, when the total 
system evolves under the density matrix dynamics of Eqs.~(2) and (3a).  

\bigskip
\centerline{\bf 4.~~Mean Field Approximation for a System}
\centerline{\bf Weakly Coupled to its Environment}
\bigskip

Let us now consider two subsystems with disjoint variables that are weakly 
coupled through an interaction term $\Delta H$ in the Hamiltonian, so that 
the total Hamiltonian appearing in Eq.~(10) is $H=H_1+H_2+\Delta H$.  We  
shall take subsystem 1 to be a measuring apparatus 
(including the microscopic system being measured), whose reduction dynamics 
we wish to follow, while we take subsystem 2 to be the external environment 
with which this measuring apparatus interacts.  We shall 
derive a mean field approximation to the dynamics, in which 
each subsystem obeys an independent system stochastic equation with a 
modified Hamiltonian, that reflects the mean interaction with the 
other subsystem.  To this end, we substitute the independent subsystem 
Ansatz $\rho=\rho_1\rho_2$ into Eq.~(10), and take the partial 
trace ${\rm Tr_2}$ to average over the subsystem 2 dynamics, giving an 
effective equation for subsystem 1, and similarly, with the roles of 1 and    
2 interchanged, to get an effective equation for subsystem 2.  
We shall assume that in the limit 
of vanishing coupling $\Delta H$,  the environment subsystem 2 is in one  
of the ensembles constructed in Sec. 3 that is a function solely of $H_2$, 
so that in the presence of $\Delta H$ we have $[\rho_2,H_2]
=O(\Delta H)$.  
We do not make a corresponding 
assumption for subsystem 1, since we will be interested in the case in 
which this is initially in a generic pure state.    

We proceed with this calculation term by term.  From the left hand side 
of Eq.~(10), substituting Eq.~(6a) we get 
$${\rm Tr}_2 d\rho={\rm Tr}_2\rho_2 d\rho_1 
+ (\rho_1 +d \rho_1){\rm Tr}_2 d\rho_2 = d\rho_1~~~,\eqno(14)$$ 
where we have used the condition  ${\rm Tr}_2 \rho_2= 1~$  which implies 
that ${\rm Tr}_2 d\rho_2=0$.  From the first term on the right hand 
side of Eq.~(10) we get 
$${\rm Tr}_2 (-i)[H,\rho]dt=-i {\rm Tr}_2 \rho_2 [H_1,\rho_1]
-i {\rm Tr}_2 [\Delta H,\rho_1\rho_2] -i{\rm Tr}_2 [H_2,\rho_2] \rho_1
~~~.\eqno(15a)$$
The first term on the right of Eq.~(15a) gives simply
$$-i[H_1,\rho_1] dt~~~.\eqno(15b)$$
Since ${\rm Tr}_2 \Delta H \rho_2={\rm Tr}_2 \rho_2 \Delta H$, the second 
term on the right of Eq.~(15a) becomes 
$$-i[{\rm Tr}_2 \rho_2 \Delta H,\rho_1]dt~~~,\eqno(15c)$$
and the third term on the right of Eq.~(15a) vanishes.  So in sum, the first  
term on the right hand side of Eq.~(10) gives 
$$-i[H_1+{\rm Tr}_2 \rho_2 \Delta H, \rho_1] dt~~~.\eqno(16)$$

We turn next to the second term on the right hand side of Eq.~(10), which 
gives $-{1\over 8}\sigma^2 dt$ times the partial trace of the 
double commutator, 
$$\eqalign{
~&{\rm Tr}_2 [H,[H,\rho]]={\rm Tr}_2 [H_1+H_2+\Delta H,[H_1+H_2+\Delta H, 
\rho_1\rho_2]]\cr
=&{\rm Tr}_2 [H_1+\Delta H,[H_1+H_2+\Delta H,\rho_1\rho_2]]\cr
=&{\rm Tr}_2\{[H_1+\Delta H,[H_1,\rho_1]\rho_2] 
+[H_1,[H_2,\rho_2]\rho_1]]+[H_1,[\Delta H,\rho_1\rho_2]] \cr
+&[\Delta H,[H_2,\rho_2]\rho_1]]+[\Delta H,[\Delta H,\rho_1\rho_2]]\}\cr 
=&[H_1+{\rm Tr}_2 \rho_2 \Delta H,[H_1+{\rm Tr}_2 \rho_2 \Delta H,\rho_1]]
+O((\Delta H)^2)~~~,\cr
}\eqno(17)$$
where we have used the facts that (i) ${\rm Tr}_2 [H_2,g(1,2)]=0$ for any 
function $g$ of variables 1,2, and that (ii) by our equilibrium 
assumption for 
the environment, $[H_2,\rho_2]$ is of order $\Delta H$. [Step (ii) 
is the only one which does not go through in the corresponding 
effective equation calculation for the environment subsystem 2, leading  
to an additional term in its effective equation of motion  given 
in Eq.~(20a) below.]

{}Finally, we turn to the third term on the right hand side of Eq.~(10), 
which gives ${1\over 2} \sigma dW_t$ times 
$$\eqalign{
~&{\rm Tr}_2[\{\rho,H\}-2\rho {\rm Tr}\rho H]
={\rm Tr}_2[\{\rho_1\rho_2,H_1+H_2+\Delta H\}-2\rho_1\rho_2{\rm Tr}_1
{\rm Tr}_2\rho_1\rho_2(H_1+H_2+\Delta H)] \cr
=&{\rm Tr}_2[\{\rho_1\rho_2,H_1+\Delta H\} +\rho_1\{\rho_2,H_2\}
-2\rho_1\rho_2{\rm Tr}_1\rho_1(H_1+{\rm Tr}_2 \rho_2 \Delta H)
-2\rho_1\rho_2{\rm Tr}_2 \rho_2 H_2]  \cr
=&\{\rho_1,H_1+{\rm Tr}_2 \rho_2 \Delta H\}
-2\rho_1    {\rm Tr}_1\rho_1(H_1+{\rm Tr}_2 \rho_2 \Delta H)~~~,\cr
}\eqno(18)$$
where no approximations have been made. 

Putting everything together, we see that the mean field approximation for 
the ``measurement'' subsystem 1 is 
$$d\rho_1=-i[H_1^{\prime},\rho_1]dt -{1\over 8}\sigma^2[H_1^{\prime},
[H_1^{\prime},\rho_1]]dt +{1\over 2}\sigma [\{\rho_1,H_1^{\prime}\} 
-2\rho_1{\rm Tr}_1\rho_1 H_1^{\prime}] dW_t+O(\sigma^2 
(\Delta H)^2 dt)~~~,\eqno(19a)$$
with the effective Hamiltonian 
$$H_1^{\prime}=H_1+{\rm Tr}_2 \rho_2 \Delta H~~~.\eqno(19b)$$
The corresponding equation for the  ``environment'' subsystem 2 is obtained 
by interchanging the labels 1 and 2, and restoring the term dropped in 
step (ii) leading to Eq.~(17), giving 
$$\eqalign{
d\rho_2=&-i[H_2^{\prime},\rho_2]dt -{1\over 8}\sigma^2[H_2^{\prime},
[H_2^{\prime},\rho_2]]dt +{1\over 2}\sigma [\{\rho_2,H_2^{\prime}\} 
-2\rho_2{\rm Tr}_2\rho_2 H_2^{\prime}] dW_t\cr
-&{1\over 8} \sigma^2 [{\rm Tr_1}(\Delta H [H_1^{\prime},\rho_1]),\rho_2]dt
+O(\sigma^2 (\Delta H))^2 dt)~~~,\cr
}\eqno(20a)$$
with the effective Hamiltonian 
$$H_2^{\prime}=H_2+{\rm Tr}_1 \rho_1 \Delta H~~~.\eqno(20b)$$
The added term on the second line of Eq.~(20a) vanishes 
through order $(\Delta H)^2$ when the reduction process 
for subsystem 1 has concluded, since then the density matrix for subsystem 
1 obeys $[H_1^{\prime},\rho_1]=0$ up to error terms of order $(\Delta H)^2$.   
As a consistency check on the calculation, we see that the mean field  
evolution equations obey ${\rm Tr_1}d\rho_1={\rm Tr_2}d\rho_2=0$, and so 
preserve the trace conditions ${\rm Tr}_1 \rho_1={\rm Tr}_2 \rho_2=1$.

\bigskip
\centerline{\bf 5.~~Dynamics of the Measurement Process:}  
\centerline{\bf When is 
Coherence Maintained, When Does the State Vector Reduce?}
\bigskip

Let us now examine the implications of Eqs.~(1-3) for measurements. 
We first have to specify the value of the parameter $\sigma$ 
governing the magnitude of the stochastic process.  If quantum 
mechanics is modified at all, it seems likely that such modifications 
come from new physics at the Planck scale, and so we adopt for this 
discussion the estimate [7, 12] $\sigma \sim M_{\rm Planck}^{-{1 \over 2}}$, 
with $M_{\rm Planck}$ the Planck mass (in units with $\hbar=c=1$).   
With this estimate, the reduction time $t_R$ in seconds for a state 
with initial energy variance $\Delta E$ is given [7,8,9] by 
$$t_R \sim \left({2.8 {\rm MeV}\over \Delta E}\right)^2~~~.\eqno(21)$$
Thus, for $\Delta E$ equal to a proton mass, $t_R \sim 10^{-5} {\rm sec}$, 
while for $\Delta E$ equal to the mass of a nitrogen molecule, 
one has $t_R \sim 10^{-8} {\rm sec}$.  
\medskip
\centerline{\bf5A.~~~Maintenance of Coherence}
\medskip
In order for stochastic energy-driven state vector reduction to give a 
viable phenomenology, it must satisfy the twin constraints of predicting 
the maintenance of coherence when this is observed, while 
predicting a rapid enough state vector reduction when  a  
probabilistic choice between alternative outcomes is observed.  We first 
discuss the constraints imposed by the maintenance of coherence.  
We begin by noting that according to Eq.~(21), the sole criterion 
governing how rapidly the state vector reduces is the energy variance; 
whether the system is microscopic or macroscopic plays no role.  
Coherent superpositions of macroscopic states, involving large numbers 
of particles, will persist in time if the energy spread between the 
superimposed states is small enough.  As a first example, consider the 
recent superconducting quantum interference device (SQUID) experiments  
[13, 14] observing the existence of coherent superpositions of macroscopic 
states consisting of oppositely circulating supercurrents.  Taking for 
discussion the experiment [13] (which of the two has 
the larger energy variance between 
the superimposed states), the energy spread $\Delta E$ is roughly 
$8.6 \times 10^{-6} {\rm eV}$, and the circulating currents each correspond 
to the collective motion of $\sim 10^9$ Cooper pairs.  According to Eq.~(21), 
despite the macroscopic structure of the state vector, 
the state vector reduction time $t_R$ for this experiment should be about 
$10^{23}~{\rm sec} \sim 3 \times 10^{15}~{\rm years}$, and so maintenance of 
coherence is expected.  

As our next example of the maintenance of coherence in macroscopic 
systems, we consider a recent experiment [15] demonstrating diffraction 
of the fullerenes $C_{60}$ and $C_{70}$.  We begin by noting that 
a diffraction pattern can be observed in a monoenergetic beam (in fact, 
this is the ideal condition for the experiment), so this class of 
experiments provides no evidence for coherent superpositions of states 
of differing energies.  However, in a realistic experiment there will be 
an energy spread in the wave packet for each particle 
constituting the beam.  To see a 
diffraction pattern, the spread in de Broglie wavelengths $\Delta \lambda$ 
should be considerably smaller than $\lambda$; adopting the very weak bound 
$\Delta \lambda \leq \lambda$, we get the requirement that the spread in 
beam momenta $\Delta p$ in each wave packet should obey $\Delta p \leq p$.  
This implies that each wave packet must have an energy spread $\Delta E$  
obeying 
$\Delta E \leq 2E_{\rm kinetic}$.  In the experiments of Ref. [15] the 
beam was obtained from an oven at approximately $900K$, and so the 
the bound on the energy spread becomes $\Delta E \leq 2 \times (3/2) 
\times 900K \sim .23 {\rm eV}$.  The corresponding state vector reduction 
time predicted by Eq.~(21) is of order $1.5 \times 10^{14}~{\rm sec} 
\sim 5\times 
10^6~{\rm years}$, and so energy driven state vector reduction plays no 
role in this experiment.  Similar estimates, and the same conclusion, would 
hold if larger objects, such as viruses, were diffracted.  

These estimates suggest that in order to try to see the 
breakdown of coherence predicted by Eq.~(21), one should consider experiments 
with systems having long lived metastable states separated by a large energy 
gap from the ground state.  In atomic systems, the requirements on stability 
of the metastable state are very severe, since for a typical atomic energy 
splitting  of a few eV, 
Eq.~(21) predicts a state vector reduction time of order 
$10^{12}~{\rm sec} \sim 3\times 10^4~{\rm years}$.  For example, in the 
quantum intermittency experiments discussed in  [16,17], the 
metastable state lifetime 
is of order 1 sec, and so stochastic state vector reduction effects are 
negligible.  A potentially more promising case is provided by certain 
long-lived nuclear isomers [18],  which are rendered metastable by their 
high spins, and which have large energy gaps from their ground states. 
{}For example, ${}^{178}{\rm Hf}$ has an isomer with a half life of 31 years 
suspended 2.4 MeV above its ground state.  Quantum mechanics predicts 
that a coherent superposition of the isomeric state and the ground state 
should be stable for time intervals that are short relative to 31 years, 
whereas Eq.~(21) predicts a spontaneous 
reduction of such a superposition to either the isomeric state or the 
ground state, with a reduction time of order 1 sec.  The only nuclear 
isomer to exist naturally on earth is the metastable isomer of 
${}^{180}{\rm Ta}$, which has a half-life of more than $10^{15}$ years, an 
energy gap of $75 {\rm keV}$ from the ground state, and which accounts 
for roughly 1 part in $10^4$ of naturally occurring tantalum.  According 
to Eq.~(21), a coherent superposition of the ground state and metastable 
isomer of ${}^{180}{\rm Ta}$ should spontaneously reduce to either 
the isomeric  
state or the ground state, with a reduction time of order 23 minutes.  
Maintenance of coherence of such a superposition for times significantly 
longer than this 
would decisively rule out Eqs.~(1-3) as a phenomenology for state vector 
reduction.  For example, if a laser using isomeric ${}^{180}{\rm Ta}$ could 
be constructed, and if the characteristic relaxation times for conventional 
sources of dissipation could be made much longer than 23 minutes, then 
the effects of Eq.~(21) might appear as an additional, unconventional  
source of stochastic fluctuations or of dissipation.    
It would clearly be of interest to work out the 
detailed implications of Eqs.~(1-3) for laser action in such a system.  

\medskip
\centerline{\bf5B.~~~Reduction in Measurement Situations} 
\medskip
We turn now to the second requirement that must be satisfied by a 
phenomenology of state vector reduction, which is 
that it should lead to rapid reduction in experimental situations where a 
probabilistic outcome is observed.  According to the von Neumann model 
for measurement [19], a measurement sets up a  correlation  
between states $|f_{\ell}\rangle$ of a quantum  system being measured, and 
macroscopically distinguishable states $|{\cal M}_{\ell}\rangle$ of the 
measuring apparatus ${\cal M}$, in such a way that an initial state  
$$|f\rangle |{\cal M}_{\rm initial}\rangle=
\sum_{\ell}c_{\ell}|f_{\ell}\rangle
|{\cal M}_{\rm initial}\rangle~~~\eqno(22a)$$ 
evolves unitarily to 
$$\sum_{\ell} c_{\ell} |f_{\ell}\rangle  |{\cal M}_{\ell}\rangle~~~.
\eqno(22b)$$
An objective state vector reduction model must then account for the selection 
of {\it one} of the alternatives 
$|f_{\ell}\rangle  |{\cal M}_{\ell}\rangle$ from this superposition, 
with a probability given by $|c_{\ell}|^2$.  
If the energy spread among the states $|f_{\ell}\rangle$ is a typical 
atomic magnitude of a few eV, then as we have seen above using Eq.~(21),  
the energy driven model of Eqs.~(1-3) cannot quantitatively account 
for state vector reduction, unless the energy spreads among the 
alternative apparatus states in the superposition are much larger.  Since 
in the ideal measurement model there is no energy transfer from the 
microscopic system to the apparatus, such an energy spread in the 
measurement apparatus states can only be present if induced by 
environmental interactions, which are ignored in the von Neumann analysis.  
If these environmentally induced energy fluctuations are large enough for the 
state vector to reduce in a time much smaller than the measurement time, then 
the observed results will agree with the Copenhagen interpretation 
of the measurement process.  If the reduction time were of the order of 
or larger than the measurement time, then Eqs.~(1-3) would predict  
stochastic fluctuations among alternative measurement outcomes, lasting 
until one is finally selected in a time roughly equal to $t_R$.    
However, as long as the apparatus states 
$|{\cal M}_{\ell}\rangle$ are orthogonal for different $\ell$, no 
quantum interferences between different outcomes are possible.   

To reiterate, for environmental interactions 
to be effective in producing state 
vector reduction, they must lead to energy fluctuations $\Delta E$ 
of the apparatus  
in the course of a measurement, that are large enough for Eq.~(21) to 
predict a reduction time $t_R$ that is less than the time it takes 
to make the measurement.  Although different measuring devices have different 
response times, we shall for assume for purposes of our discussion that 
relevant measurement times range down to $10^{-8}$ seconds,   
which requires for reduction a $\Delta E$ ranging up to $\sim 30$GeV. 
We shall consider three possible sources of 
energy fluctuations:  thermal energy fluctuations, fluctuations in apparatus 
mass from particle accretion processes, and fluctuations in 
apparatus mass from amplified fluctuations in the currents that 
actuate the indicator devices.  

Thermal energy and temperature fluctuations in a canonical ensemble, that is, 
with fixed particle number, are governed by the  equations 
$$ \langle (\Delta E)^2 \rangle_{AV}=k_B T^2 C_V~,~~~
\langle (\Delta T)^2 \rangle_{AV}=k_B T^2/C_V~~~,\eqno(23)$$
with $k_B$ Boltzmann's constant and with $C_V$ the heat capacity. From these 
formulas, and the values of the heat capacity and thermal conductivity for 
various substances, together with the formulas governing the surface  
radiation rate, one can estimate that when a body is large enough for 
the energy fluctuations at room temperature 
to be of order 1 to 30 GeV, the thermal relaxation time over which such 
energy fluctuations occur is much larger than measurement times of interest.  
(For example, for 1 gm of water at room temperature, 
the root mean square energy fluctuation is 
$\sim 14 {\rm GeV}$, and the thermal conduction 
relaxation time is $>200$  sec.)    
The reason for this is that the rate for 
heat transfer processes is proportional to the temperature difference  
$\Delta T$, and Eq.~(23) shows that when a body is large, the temperature 
fluctuation $\Delta T$ is small.  
Thus we find that thermal energy fluctuations do not in general 
obey the criterion   
stated above for relevance to state vector reduction, that the energy    
fluctuation should occur within the measurement time. 

A more significant source of energy fluctuations comes [7] from particle 
accretion processes, for which we formulate a simple model within the 
mean field framework of Sec. 4.  
Consider a measuring apparatus which has $N$ surface accretion sites for 
molecules of mass $m$.  In Fock space representation, its Hamiltonian can 
be written as 
$$H_1=H_0+\sum_{j=1}^N m a_j^{\dagger} a_j~~~,\eqno(24a)$$
with $H_0$ the bulk Hamiltonian for the apparatus, and with $a_j^{\dagger}$  
and $a_j$ respectively the creation and  annihilation operators
for the accreted molecules.  We assume the environment to contain a large 
number M of molecules that can be accreted onto the surface, with creation  
and annihilation operators $b_k^{\dagger},b_k~~,k=1,...,M,$ and with a   
coupling to the accretion sites given by 
$$\Delta H=\sum_{j=1}^N\sum_{k=1}^M [A_{jk} a_j^{\dagger}b_k +{\rm adjoint}]
~~~.\eqno(24b)$$
This interaction Hamiltonian conserves the total number operator 
$$N=\sum_{j=1}^N a_j^{\dagger}a_j + \sum_{k=1}^M b_k^{\dagger} b_k ~~~,
\eqno(25)$$
in other words, the total number of molecules accreted onto the surface or 
remaining in the environment is constant.  

In typical measurement situations, the environment density matrix will 
be diagonal in the number operator $\sum_{k=1}^M b_k^{\dagger} b_k$ of the 
molecules being accreted.  In this case, which we term ``incoherent'', 
the environmental expectation of $\Delta H$ vanishes, 
$${\rm Tr}_2 \rho_2 \Delta H =0~~~,\eqno(26)$$ 
and the reduction process is governed, according to Eqs.~(19a,b),  
by the measurement system Hamiltonian $H_1$ alone.  (For a discussion of  
the coherent case, in which the environmental expectation of $\Delta H$ is 
nonzero, see the Appendix.)  The Hamiltonian 
$\Delta H$ still plays a role, since in order $\Delta H$ in probability 
amplitudes [corresponding to order $(\Delta H)^2$ in probabilities or 
transition rates] it leads to 
a sticking probability and an evaporation probability per unit time,  
respectively, for a molecule in 
the environment to accrete to the surface of the apparatus, and for a 
molecule already accreted to evaporate.  As a result of these nonvanishing 
transition probabilities, the number of molecules accreted to the surface 
is constantly fluctuating.  Assuming a simple colloid 
statistics model [20] in  
which each accretion site can hold only one molecule, the number of accreted 
molecules $n$ obeys a Poisson distribution 
$\sigma(n,X)=e^{-X}X^n/n!$ with the mean $X$ proportional to the sticking 
probability and inversely proportional to the evaporation rate, and with the 
the root mean square fluctuation in the number of accreted molecules 
equal the square root of the mean number $X$ of accreted molecules.  

Since distinguishable measurement outcomes must involve different 
configurations of the apparatus with respect to its environment,  
they will have  
different values of the accretion numbers $a_j^{\dagger}  a_j$ associated 
with the $N$ accretion sites.  Thus, the energy eigenvalue $H_1$ of the 
measurement apparatus will differ for each distinguishable measurement 
outcome,  with the spread of eigenvalues between any two outcomes being 
typically the mass of the accreted molecules $m$ times the root mean 
square fluctuation in the number of accreted molecules. This statement  
assumes that the  
flux of accreting molecules in the environment is high enough for such a  
fluctuation to actually occur during the state vector 
reduction time.  In estimating when   
this condition holds, we will follow the review of Redhead [21] in 
assuming that the sticking probability is of order unity, in which 
case the minimum time for one molecule to be accreted onto an area of   
1 ${\rm cm}^2$ 
can be read of  from the molecular flux versus pressure 
tabulated in Table 2 of [21].  At room temperature and 
atmospheric pressure (760 Torr)  
the time for one molecule to be accreted onto an area of 
$1 {\rm cm}^2$ is $3\times 10^{-24}~{\rm sec}$, while
at an ultrahigh vacuum of $10^{-13}$ Torr it is 
$3 \times 10^{-8} ~{\rm sec}$. 
Thus, for an apparatus in the atmosphere at standard temperature and  
pressure, 
where the bulk of the accreting atoms are nitrogen molecules, the minimum 
apparatus area required for one molecule to accrete in a reduction time 
of $10^{-8} {\rm sec}$ (corresponding to a  $\Delta E$ equal to the mass of 
a nitrogen molecule) is $3 \times 10^{-16} {\rm cm}^2$, with the   
corresponding minimum area needed at a pressure of $10^{-13}$ Torr equal to 
$3 {\rm cm}^2$.  

According to [21], the nighttime pressure at the surface 
of the moon is about $10^{-13}$ Torr, while the pressure in interstellar 
space (within the galaxy) has been estimated as $10^{-18}$ Torr.  
Under the assumption of a
sticking probability of order unity, the mass accretion rate scales as 
the pressure divided by the mean molecular velocity.  While molecular 
velocities away from the vicinity of the earth vary over a wide  
range, with effective temperatures in interstellar space ranging [22] from 
typically 50--100 to $10^6$ degrees Kelvin, 
we can get an estimate that is high by at most a factor of 2 or 3  by    
neglecting the velocity factor, and simply assuming that the mass accretion 
rate scales with the pressure from the values given in the table of [21].  
With this assumption, the minimum apparatus area needed for a reduction time  
of $10^{-8} {\rm sec}$ is $3 {\rm cm}^2$ at the surface of the moon, and 
is  $3  \times 10^5 {\rm cm}^2 = 30 {\rm m}^2$ in interstellar space.   
In intergalactic space, the predominant matter [22] is highly 
ionized hydrogen, with an effective temperature of order $10^4$ degrees 
Kelvin and a density of $\sim .23$ proton per cubic meter.  
In this environment, 
the minimum apparatus area needed for a reduction time of $10^{-8} {\rm sec}$ 
is around $8 \times 10^5 {\rm m}^2$ (corresponding to the accretion of 28 
protons).  
If we 
were only to demand reduction in $3 \times 10^{-4}$ second, then 
the needed apparatus 
area in interstellar space would decrease to less than 
$10 {\rm cm}^2$, while that in 
intergalactic space would 
decrease to $\sim 1 {\rm m}^2$.  Thus, a 
capsule large enough to sustain  
Schr\"odinger's cat, situated in intergalactic space, would have a reduction 
time stimulated by collisions with molecules in the intergalactic medium 
much smaller than the length of time needed to ascertain whether the cat 
were dead or alive!  Perhaps more to the point, in a typical high 
precision molecular beam experiment [23], the beam velocity is of order 
$10^5 {\rm cm}/{\rm sec}$, and the beam length is of order 2.7 meters.  
Hence the time for the beam to traverse the apparatus is $2.7 \times 
10^{-3} {\rm sec}$, and so the reduction time in intergalactic space 
for a capsule large enough 
to enclose the apparatus would be smaller, by at least an order  
of magnitude, than the measurement time.  Clearly, in this situation  
the limits predicted by Eq.~(21) are being pushed, and there could 
be realizable experiments which, in intergalactic space, would be predicted 
to start to show evidence of the stochastic fluctuation between outcomes 
characteristic of the time evolution of the state vector in stochastic 
reduction models.  But it seems unlikely that such an experiment could 
be devised within the confines of the solar system -- the ambient matter 
fluxes are too high.

In making some of the above estimates, it is convenient to have 
an alternative form 
of Eq.~(21) that takes into account the accretion rate limit on $\Delta E$, 
and which is derived as follows. 
Let ${\cal M}$ be the mass accretion rate 
per unit area of 
the apparatus in units ${\rm sec}^{-1}$, so that the mass accretion on area 
$A$ in  
time $t_R~ {\rm sec}$ is $\Delta E =A {\cal M} t_R$.  Substituting 
this into Eq.~(21) and solving for $t_R$ gives 
$$t_R = \left( {2.8 {\rm MeV} \over A {\cal M}}\right)^{2 \over 3}~~~.
\eqno(27)$$
This formula can be used whenever at least one molecule is accreted in the 
time $t_R$.  
Given ${\cal M}$, we can calculate the area ${\cal A}$ corresponding to 
a given reduction time, and vice versa.  For example, from Eq.~(27) 
we find that 
for an apparatus of area $1 {\rm cm}^2$ in the atmosphere at standard 
temperature and pressure, the reduction time is $t_R = 5 \times 
10^{-19} {\rm sec}$, corresponding to the accretion of $\sim 1.5 \times 10^5$   
molecules in the time $t_R$.  

Throughout this analysis, we have assumed that the Hamiltonian that 
is relevant for the stochastic Schr\"odinger equation is the total 
Hamiltonian 
$$H=\int d^3x T_{00}(x)~~~\eqno(28)$$
defined by gravitational couplings to the stress-energy tensor 
$T_{\mu\nu}(x)$, which includes rest mass terms.  
Although in non-relativistic 
quantum mechanics one often drops rest mass terms when they lead to 
irrelevant constant energy shifts, there is no reason in principle to 
do so.  In fact, in the standard model of elementary particles, all fermion 
rest masses arise from the Yukawa couplings of the fermions to the Higgs 
particle, so that from this point of view rest masses are not an 
additive constant term in the Hamiltonian, but are a dynamical product of 
interactions.   We have also assumed that the relevant surface area is 
that of the whole apparatus, rather than just that of components of 
potentially small area such as solid state detectors, emulsions, or 
particle collector cups.  This assumption is motivated by our decoupling 
analysis of Sec. 2, where we saw that only noninteracting systems can 
be assumed (under certain equilibrium conditions) to decouple.  The  
components of an apparatus (power supplies, magnets, vacuum pumps, detectors,  
indicator pointers, magnetic recording domains) 
are not in equilibrium and are in interaction with one another, and so 
using the area of the whole apparatus, rather than of just the smallest 
components, seems justified.   

We turn finally to a third potential source of energy fluctuations, arising 
from the amplified fluctuations in the currents which actuate experimental 
indicating or recording devices.  Of course, if power sources are included, 
there are no overall current fluctuations, but power supplies are typically 
large in area and so when included in the system the 
accretion analysis just given   
indicates rapid reduction times.   In a typical electrically amplified  
measurement, a final total charge transfer $Ne$ (with $e$ the charge of 
an electron) actuates an 
indicator or recording device. 
Assuming that the fluctuation in the current is the amplified fluctuation 
in the initially detected signal, for amplification gain $G$ we have 
$\Delta N \sim G \times (N/G)^{1\over 2} =(NG)^{1 \over 2}$, an estimate 
which agrees within factors of order unity with the standard noise estimate 
for photomultipliers [24].  Let us take  $N$ to correspond to a charge 
transfer of 1 milliampere (a  voltage change of 10 volts at 10 k$\Omega$ 
impedance) 
over a $10^{-8}$ second pulse , so that   
$N \sim 6 \times 10^7$, and assume a gain $G \sim 
10^4$, giving $\Delta N \sim 8 \times 10^5$.  Multiplying by the 
electron mass of 
$ .5 \times 10^{-3} {\rm GeV}$, we find that the corresponding energy  
fluctuation is $\Delta E \sim 4 \times 10^2 {\rm GeV}$, which leads to 
state vector reduction in $5 \times 10^{-11} {\rm sec}$.  Thus, electric 
current fluctuations play a significant role in state vector reduction 
when the ``apparatus'' is defined to exclude power sources.  

Our overall conclusion is that conditions under which laboratory 
experiments are performed, as well as conditions under which space 
capsule experiments might be performed 
in the foreseeable future, are consistent 
with state vector reduction times as estimated by Eq.~(21) that are well 
within experimental measurement times.  

\medskip
\centerline{\bf5C.~~~Experiments with Semi-Silvered Mirrors} 
\medskip
In the preceding two subsections we have considered the case in which the 
energy variance is so small that coherence is maintained, and the case 
in which the energy variance is large enough that reduction proceeds 
rapidly to completion.  Let us now briefly consider a case that contains 
elements of both, in which an apparatus is constructed using semi-silvered 
mirrors (for photons) or thin detectors (for particles), so that there 
is a probability amplitude $\alpha$ for no interaction with the 
apparatus and the maintenance of coherence, and a probability amplitude  
$\beta$ 
for a measurement to take place.  The state vector then has the form, 
after the measurement interaction, 
$$\alpha |f\rangle |{\cal M}_{\rm initial}\rangle+ \beta
\sum_{\ell} c_{\ell} |f_{\ell}\rangle  |{\cal M}_{\ell}\rangle~~~.
\eqno(29)$$
Assuming that $|{\cal M}_{\rm initial}\rangle$ and $|{\cal M}_{\ell}\rangle$
differ sufficiently in energy for reduction to take place, there are now 
two possible classes of outcomes.  With probability $|\beta c_{\ell}|^2$ 
the $\ell$th measurement outcome is observed, while with probability 
$|\alpha|^2$ the initial state is unchanged, corresponding to   
transmission through the semi-silvered mirror or thin detector.  
In making the latter 
assertion, we are using the fact that, as shown in Ref. [9], the model 
of Eqs.~(1-3) obeys the L\"uders projection postulate.  That is, a    
component of the wave function lying within a submanifold of Hilbert space 
that is 
energy degenerate (or nearly degenerate, in the sense of Sec. 5A) survives  
unchanged in form, with the appropriate probability, as an outcome of the 
reduction process.  This corresponds exactly to what happens when a beam 
is transmitted through a partially silvered mirror or a thin detector.  This 
discussion generalizes immediately to the case in which the transmitted 
beam has different phase shifts in the various terms in the superposition 
over wave function components $|f_{\ell}\rangle$.  

\bigskip
\centerline{\bf 6.~~Coexistence of Reduction and Decoherence}
\bigskip
In the previous section, we considered the effects on an 
apparatus of inelastic collisions, in which the mass of the apparatus 
fluctuates.  However, an apparatus considerably more 
frequently suffers elastic 
collisions with atoms and photons in its environment, which are responsible 
for the decoherence effects [25,26,27] that have been much discussed in 
the literature.  We shall argue in this section that decoherence effects 
do not substantially modify the results of the preceding section. 

We begin with a general argument that is independent of the details of 
modeling decoherence.  When elastic interactions with the environment are 
taken into account, the effective apparatus wave function has to be 
extended to include the wave functions of all particles with which it 
interacts during the reduction time $t_R$.  
Since this extension of the apparatus definition increases its area, the 
rate of mass fluctuations is increased, and the effective reduction time 
estimated in the preceding section is, if anything, decreased.
(Because  $t_R$ is in general 
shorter than the time for molecules or photons in the environment to 
collide with one another, we do not have to continue this enlargement
of the apparatus another step to include the particles with which the 
decohering particles interact -- they can be regarded as effectively 
noninteracting.  For example, we estimated above that in air at standard
temperature and pressure, the reduction time $t_R$ for an apparatus of area
$1 {\rm cm}^2$ is $\sim 5\times 10^{-19} {\rm sec}$ (during which time 
it accretes $\sim 1.5 \times 10^5$ air molecules), whereas the mean 
time between collisions of air molecules with each other 
is $\sim 10^{-10} {\rm sec}$.
Since the reduction time through accretion 
scales as the inverse ${2\over 3}$ power of the density of the environmental  
medium, while the time between collisions of a molecule scales inversely 
as the density, in more dilute environments this inequality 
gets stronger.)   With the definition of the apparatus wave function 
extended in this way, Eq.~(22b) is modified to read 
$$\sum_{\ell} c_{\ell} |f_{\ell}\rangle  |{\cal M}_{\ell}\rangle 
|\Psi_{e\ell}\rangle~~~,\eqno(30)$$
with $|\Psi_{e\ell}\rangle$ the environmental 
wave function associated with the 
$\ell$th apparatus state. We can now apply the analysis developed above, 
using Eqs.~(1-3) to describe the stochastic reduction of the wave function of 
Eq.~(30), 
with the conclusions reached above unaltered.  The wave function of the 
extended apparatus remains unit normalized, and its density matrix 
remains a pure state density matrix, which stochastically evolves to   
an energy eigenstate with the rate given by Eq.~(21).  

Decoherence effects manifest themselves by the exponential decay with time 
of the  inner product $\langle \Psi_{e\ell}|\Psi_{e \ell^{\prime}}\rangle $,  
for $\ell \neq \ell^{\prime}$, 
so that the environmental states associated with different measurement 
outcomes become rapidly orthogonal.  Correspondingly,  
in the reduced density matrix for the original unextended apparatus, 
obtained by tracing out the environmental degrees of freedom, there is an  
exponential decay of the off diagonal matrix elements.   
{}For the superposition of energy eigenstates 
that is relevant for our discussion, the relevant decay or decoherence 
rate of the off diagonal reduced matrix element is given by [25,26] 
$$D =N_{\rm scatt} {\rm Re}[1-
\langle S_{\ell}^{\dagger} S_{\ell^{\prime}} \rangle]~~~.\eqno(31a)$$ 
Here $N_{\rm scatt}$ is the number of scatterings by 
environmental particles in 
unit time,   
$\langle ...\rangle$ is an expectation in the state of the scattering 
particle, and  $S_{\ell}$ and $S_{\ell^{\prime}}$ are the 
scattering matrices acting on this 
particle when it scatters on the respective  
components of the apparatus wave function with state labels $\ell$ and 
$\ell^{\prime}$.  In our context, these apparatus states differ only by 
the addition of some number of accreted molecules, and so in a 
weak scattering 
approximation the product $S_{\ell}^{\dagger} S_{\ell^{\prime}}$ can be 
approximated as $S_M$, with $S_M$ the scattering matrix for an environmental 
particle to scatter from the accreted molecules.  In this approximation, 
Eq.~(31) simplifies to 
$$D =N_{\rm scatt} {\rm Re}[1-
\langle S_M \rangle]~~~.\eqno(31b)$$ 
Expressing the scattering matrix $S_M$ in terms of the corresponding 
scattering amplitude, and using the optical theorem, Eq.~(31b) reduces [25] 
to 
$$D={1\over 2} \times {\rm Scattering~Rate}~~~.\eqno(32)$$
Combining Eq.~(32) with our estimates above for an area $1{\rm cm}^2$ 
apparatus in air, 
the decoherence rate arising from molecules of the environment scattering on 
the molecules accreted during the reduction time is ${1\over 2} \times 
10^{10} {\rm sec}^{-1} \times 1.5 \times 10^5$$\sim .7 \times 10^{15}
{\rm sec}^{-1}$,  several 
orders of magnitude smaller than the reduction rate $t_R^{-1}\sim 
2 \times 10^{18}{\rm sec}^{-1}$, and so 
decoherence effects are in fact unimportant over the duration of the 
reduction process.  
Since the ratio of the reduction rate to the decoherence rate (calculated 
for the number of molecules accreted over the reduction time) scales with 
area as ${\cal A}^{1\over 3}$, this conclusion remains true down to an 
apparatus area of $\sim 4 \times 10^{-11} {\rm cm}^2$, where the two rates 
become approximately equal.  Thus, even for a very small apparatus 
the environmental states $|\Psi_{e\ell}\rangle$ 
are nearly identical to the initial environmental state $|\Psi_e\rangle$, and so 
the wave function of Eq.~(30) is negligibly entangled with the environment. 
Therefore instead of using the extended apparatus to discuss the 
reduction process, one is justified in ignoring decoherence and using the 
original unextended apparatus as in Eq.~(22b).  Our conclusion here for 
energy driven reduction models differs significantly from that reached 
[28,29] for spontaneous localization models, where decoherence effects over 
the reduction time are substantial; however, in these models the
various apparatus states in the superposition of Eq.~(22b)  
differ by a displacement of the center of mass of some part of the 
apparatus, which gives a decoherence    
effect proportional to the (macroscopic) scattering cross section of that 
part of the apparatus which is displaced.  

\bigskip
\centerline{\bf 7.~~ Discussion and Conclusions}
\bigskip

The analysis we have given of a number of aspects of the  
effect of the environment on the measurement process, including the   
decoupling of isolated systems from environments in equilibrium, 
the effect of energy fluctuations 
induced by mass accretion, and the effect of decoherence processes, 
supports the view 
that the energy-driven stochastic Schr\"odinger equation gives a viable 
phenomenology of state vector reduction.   According to this picture, 
a measurement takes place when the different outcomes are characterized by 
sufficiently large environmentally induced energy fluctuations in the 
apparatus for the state vector reduction process, which is driven by the 
energy variance, to proceed rapidly to completion.    
The infinite Von Neumann regression (of an apparatus measuring an apparatus  
measuring an apparatus..., ad infinitum) terminates, when the apparatus 
size is large enough for its 
energy fluctuations to lead to state vector reduction within the specified 
observation time.  
This requirement on apparatus size meshes 
in a natural way with the intuitively obvious requirement 
that in a measurement, 
different experimental outcomes must be macroscopically distinguishable.

\bigskip
\centerline{\bf Acknowledgments}
\bigskip
I wish to thank Jeeva Anandan, John Bahcall, 
Angelo Bassi, Todd Brun, Sudip Chakravarty, 
GianCarlo Ghirardi, Siyuan Han, Larry Horwitz, Lane Hughston, James Lukens, 
Indrajit Mitra, 
Ian Percival, Leo Stodolsky, and Frank Wilczek for stimulating discussions 
or email correspondence, 
and to acknowledge the  
hospitality of the Aspen Center for Physics, where part of this work was 
done.  
This work was supported in part by the Department of Energy under
Grant \#DE--FG02--90ER40542.
\vfill\eject
\bigskip
\centerline{\bf Appendix.~~~Coherent Case of the Accretion Model}  
\bigskip

Anandan [30] has raised the interesting question of 
whether state vector reduction 
can proceed to a coherent state endpoint.  One way in which this can 
happen is when the signal amplification process involves
coherent states,  as discussed in [27].  Another way, 
which we shall discuss here, corresponds to the ``coherent'' case of 
the accretion model formulated in Sec.~5B, in which the environment 
is in a coherent state, so that the environmental expectation of $\Delta H$ 
is nonzero.  Assuming for simplicity that there is only one accretion site, 
which can be multiply occupied, 
we have then 
$$H_1+{\rm Tr}_2 \rho_2 \Delta H =H_0 +ma_1^{\dagger}a_1 + 
\lambda a_1^{\dagger} 
+\lambda^* a_1~~~,\eqno(A1a)$$
with $\lambda$ given by 
$$\lambda=\sum_{k=1}^M A_{1k} {\rm Tr}_2 \rho_2 b_k~~~.\eqno(A1b)$$

Assuming $H_0$ to commute with $a_1$, Eqs. (A1a) and (A1b) describe 
the zero forcing frequency limit of the 
forced harmonic oscillator, which has been extensively 
studied [31,33], and can   
be succinctly solved by coherent state methods [32,33]. 
Defining $z$  and $c_1$  
by 
$$z\equiv -\lambda/m~,~~~c_1\equiv a_1-z=a_1+\lambda/m~~~,\eqno(A2a)$$
we have 
$$H_1=H_0+m c_1^\dagger c_1 +{\rm constant}~~~,\eqno(A2b)$$
which in its $c_1$ dependence is a standard harmonic oscillator.  The $c_1$ 
oscillator ground state $|0\rangle$ obeys 
$$c_1|0\rangle =0 \Rightarrow  a_1|0\rangle = z|0\rangle~~~,\eqno(A2c)$$
in other words, $|0\rangle $ is a coherent state in terms of the original 
operators $a_1$.

Ignoring an overall constant arising from terms in Eq.~(A2b) that commute 
with $a_1$, the general eigenstate of Eq.~(A2b) is $|n\rangle$, with 
$n$ the number of $c_1$ quanta, and has energy eigenvalue $mn$.
This state is a coherent superposition of states with different numbers of 
molecules on the accretion site.  For energy eigenvalue $n$, 
the probability $P(n|k)$ of finding $n-k$ molecules on the site can be 
exactly expressed [31,33] as a Laguerre polynomial, and 
for $|z|<<1$ and $n$ large can be approximated [34] as 
$$P(n|k)\simeq [J_{|k|}(2 n^{1\over 2} |z|)]^2 ~~~,\eqno(A3a)$$
with $J_k$ the order $k$ Bessel function; the Bessel function addition 
formula [35] 
$$1= J_0(w)^2 + 2 \sum_{n=1}^{\infty} J_n(w)^2 ~~~\eqno(A3b)$$ 
implies that the probabilities of Eq.~(A3a) sum to unity, 
$$\sum_{k=-\infty}^{\infty} P(n|k)=1~~~.\eqno(A3c)$$
Equation (A3a) is rapidly oscillating as a function of $k$, but using the 
asymptotic estimate [36]  
$$J_{\nu}(\nu \sec \beta) \simeq \left( 
{2\over \pi \nu \tan \beta}\right)^{1 \over 2} \cos(\nu \tan \beta -\nu \beta 
-{1\over 4} \pi)~~~,\eqno(A4a)$$
it is easily seen that the averaged envelope of $P(n|k)$ is 
given by 
$$\overline{P(n|k)}\simeq {1\over \pi} 
{1\over (4n|z|^2 -k^2)^{1\over 2} }~~~, 
\eqno(A4b)$$ 
showing that the values of $k$ are mainly 
distributed (apart from an exponentially decaying tail)
between $-2n^{1 \over 2} |z|$ and $2 n^{1\over 2} |z|$.

\vfill \eject

\bigskip
\bigskip
\centerline{\bf References}
\bigskip
\noindent
\item{[1]} Pearle P 1976 {\it Phys. Rev.} D {\bf 13} 857
\item{~~~} Pearle P 1979 {\it Int. J. Theor. Phys.} {\bf 18} 489
\bigskip 
\noindent
\item{[2]} Gisin N 1984 {\it Phys. Rev. Lett.} {\bf 52} 1657
\bigskip
\noindent
\item{[3]}  Ghirardi G C, Rimini A and Weber T 1986 {\it Phys. Rev. D} 
{\bf 34} 470
\bigskip
\noindent
\item{[4]} Ghirardi G C, Pearle P and Rimini A 1990 {\it Phys. Rev. A} 
{\bf 42} 78
\bigskip
\noindent
\item{[5]}  Di\'osi L 1988 {\it J. Phys. A: Math. Gen.} {\bf 21} 2885
\item{~~~}  Di\'osi L 1988 {\it Phys. Lett.} {\bf 129} A 419
\item{~~~}  Di\'osi L 1988 {\it Phys. Lett.} {\bf 132} A 233
\bigskip
\noindent
\item{[6]}  Gisin N 1989 {\it Helv. Phys. Acta.} {\bf 62} 363
\bigskip
\noindent
\item{[7]}  Hughston L P 1996 {\it Proc. Roy. Soc.} A {\bf 452} 953
\bigskip
\noindent
\item{[8]}  Adler S L and Horwitz L P 2000 {\it J. Math. Phys.} {\bf 41} 2485
\bigskip
\noindent
\item{[9]}  Adler S L, Brody D C, Brun T A and Hughston L P 2001 
{J. Phys. A: Math. Gen.} {\bf 34} 8795
\bigskip
\noindent
\item{[10]}  For a survey, see Pearle P 2000  {\it Open 
Systems and Measurements in Relativistic Quantum Theory (Lecture Notes in 
Physics {\bf 526})} ed H-P Breuer and F Petruccione (Berlin: Springer)  
See also citations [19]--[23] of Adler and Brun, Ref. [11].
\bigskip
\noindent
\item{[11]}  Adler S L and Brun T A 2001 {\it J. Phys. A: Math. Gen.} 
{\bf 34} 4797
\bigskip
\noindent
\item{[12]}  See work of Di\'osi, Ghirardi {\it et al}, and Penrose cited 
in Sec. 11 of Ref. [7], and also  D. I. Fivel, preprint quant-ph/9710042  
\bigskip
\noindent
\item{[13]}   Friedman J R, Patel V, Chen W, Tolpygo S K and 
Lukens J E 2000 {\it Nature} {\bf 406} 43
\bigskip
\noindent
\item{[14]}  van der Wal C H, ter Haar A C J, Wilhelm F K, 
Schouten R N, Harmans C J P M, Orlando T P, Lloyd S and  
Mooij  J E 2000 {\it Science} {\bf 290} 773  
\bigskip
\noindent
\item{[15]}  Nairz O, Brezger B, Arndt M and Zeilinger  A 2001, 
preprint quant-ph/0110012
\bigskip 
\noindent
\item{[16]}  Dehmelt H 1981 {\it Laser Spectroscopy V  
(Springer Series in Optical Sciences {\bf 3})} 
ed A R W McKellar, T Oka and 
B P Stoicheff (Berlin: Springer) 
\item{~~~~}  Dehmelt H 1982 {\it IEEE Trans. Instrum. Meas.} {\bf 2} 83
\item{~~~~}  Dehmelt H 1983 {\it Advances in Laser  
Spectroscopy  (Nato Advanced Study Institute {\bf 95})}  
ed F T Arecchi, F Strumia and H Walther (New York: Plenum)
\bigskip
\noindent
\item{[17]}  Porrati M and Putterman S 1987 {\it Phys. Rev.} A {\bf 36} 929
\item{~~~~}  Cohen-Tannoudji C and Dalibard J 1986 
{\it Europhys. Lett.} {\bf 1} 441
\item{~~~~}  Porrati M and Putterman  S 1989 {\it Phys. Rev.} A {\bf 39} 
3010
\bigskip
\noindent
\item{[18]}  Walker P and Dracoulis  G 1999 {\it Nature} {\bf 399} 35; 
I wish to thank F Wilczek for bringing this reference to my attention.
\bigskip
\noindent
\item{[19]}   von Neumann J 1932 {\it Mathematische Grundlagen der  
Quantenmechanik} Chap. VI (Berlin: Springer)   (Engl. Transl. R T Beyer 
1971 {\it Mathematical Foundations of Quantum Mechanics})(Princeton: 
Princeton University Press)  For a recent pedagogical exposition  
see Adler S L 1995 {\it Quaternionic Quantum 
Mechanics and Quantum Fields} Sec. 14.2 (New York: Oxford University Press)     
\bigskip
\noindent
\item{[20]}  Adler S L and Mitra I 2000 {\it Phys. Rev. E} {\bf 62} 4386
\bigskip
\noindent
\item{[21]}  Redhead P A 1996  {\it Macmillan Encyclopedia of 
Physics} Vol. IV p. 1657 ed J S Rigden  (New York: Simon \& Schuster 
Macmillan)
\bigskip
\noindent
\item{[22]}  Verschuur G L 1979   
{\it Encyclopedia 
Britannica, 15th ed.}  Vol. 9 p. 790  (Chicago: Britannica)
\item{~~~~} J N Bahcall, private communication.
\bigskip
\noindent
\item{[23]}  Ramsey N F 1969  {\it Molecular Beams} p. 356
(Oxford: Oxford University  Press)
\bigskip
\noindent
\item{[24]}  Preston D W and Dietz E R 1991 
{\it The Art of Experimental Physics} p. 380 (New York: John Wiley)
\bigskip
\noindent
\item{[25]}  Harris R A and Stodolsky L 1981 {\it J. Chem Phys.} 
{\bf 74} 2145
\item{~~~~}  Stodolsky L 1996 {\it Acta Phys. Polon.} {\bf 27} 1915
\item{~~~~}  Stodolsky L 1999 {\it Phys. Rep. } {\bf 320} 51
\bigskip
\noindent
\item{[26]}  Joos E and Zeh H D 1985 {\it Zeit. Phys.} B {\bf 59} 223
\bigskip 
\noindent 
\item{[27]}  Glauber R J 1986 {\it New Techniques and Ideas in 
Quantum Measurement Theory}  ({\it Annals of the New 
York Academy of Sciences} {\bf 480}) ed D M Greenberger 
(New York: New York Academy of Sciences)
\bigskip
\noindent
\item{[28]}  Tegmark M 1993 {\it Found. Phys. Lett.} {\bf 6} 571
\bigskip
\noindent
\item{[29]}  Benatti F, Ghirardi G C and Grassi R 1995 {\it Advances 
in Quantum Phenomena} ed E G Beltrametti and J-M L\'evy-Leblond 
(New York: Plenum Press)
\bigskip
\noindent
\item{[30]}  Anandan J, private communication
\bigskip
\noindent
\item{[31]}  Fuller R W, Harris S M and Slaggie E L 1963 {\it Am. J. Phys.} 
{\bf 31} 431
\bigskip
\noindent
\item{[32]}  Klauder J R and Skagerstam B-S 1985 {\it Coherent States} 
(Singapore: World Scientific)
\bigskip
\noindent
\item{[33]}  Carruthers P and Nieto M M 1965 {\it Am. J. Phys.} 
{\bf 33} 537
\bigskip
\noindent
\item{[34]}  Gradshteyn I S  and Rhyzik I M 1965 {\it Table of Integrals,       
Series, and Products}  p. 1039 Eq. 8.978-2  (New York: Academic Press)
\bigskip
\noindent
\item{[35]}  Gradshteyn I S and Rhyzik I M, Ref. [34] p. 980 Eq. 8.536-3
\bigskip
\noindent
\item{[36]}  Abramowitz M and Stegun I 1965 {\it Handbook of 
Mathematical Functions} p. 366 Eq. 9.3.3   (Washington, D. C.:  
U. S. Government Printing Office) 

\bigskip
\noindent
\bigskip
\noindent
\vfill
\eject
\bigskip
\bye